\documentclass[reprint, nofootinbib]{revtex4-2}

\usepackage{graphicx}
\usepackage{dcolumn}
\usepackage{bm}
\usepackage{float}
\usepackage{hyperref} 
\hypersetup{
	colorlinks=true,       
	linkcolor=black,        
	citecolor=black,        
	filecolor=magenta,     
	urlcolor=black         
}

\begin{document}

\preprint{APS/123-QED}

\title{Using AI Large Language Models for Grading in Education:\\A Hands-On Test for Physics}


\author{Ryan Mok}

\author{Faraaz Akhtar}

\author{Louis Clare}

\author{Christine Li}

\author{Jun Ida}

\author{Lewis Ross}

\author{Mario Campanelli}
 \altaffiliation{Corresponding author, Professor of Physics University College London}
 \email{m.campanelli@ucl.ac.uk}
 
\affiliation{%
 Department of Physics and Astronomy\\
 University College London
}%

\date{\today}%

\begin{abstract}

Grading assessments is time-consuming and prone to human bias. Students may experience delays in receiving feedback that may not be tailored to their expectations or needs.
 Harnessing AI in education can be effective for grading undergraduate physics problems, enhancing the efficiency of undergraduate-level physics learning and teaching, and helping students understand concepts with the help of a constantly available tutor. This report devises a simple empirical procedure to investigate and quantify how well large language model (LLM) based AI chatbots can grade solutions to undergraduate physics problems in Classical Mechanics, Electromagnetic Theory and Quantum Mechanics, comparing humans against AI grading. The following LLMs were tested: Gemini 1.5 Pro, GPT-4, GPT-4o and Claude 3.5 Sonnet. The results show AI grading is prone to mathematical errors and hallucinations, which render it less effective than human grading, but when given a mark scheme, there is substantial improvement in grading quality, which becomes closer to the level of human performance - promising for future AI implementation. Evidence indicates that the grading ability of LLM is correlated with its problem-solving ability. Through unsupervised clustering, it is shown that Classical Mechanics problems may be graded differently from other topics. The method developed can be applied to investigate AI grading performance in other STEM fields.


\end{abstract}

\maketitle


\section{Introduction}

\subsection{Motivation}

AI has profoundly impacted nearly every academic field and industry in recent years. Its rapid growth in public media and widespread use has been driven by the release of an ‘AI chatbot’ known as Chat Generative Pre-Trained Transformer, ChatGPT (created by OpenAI), followed swiftly by several other large language models (LLM)-based chatbots developed by companies such as Google and Meta. These chatbots, capable of engaging in text-based conversations, are built from complex large language models developed through machine learning algorithms. The recent iterations of LLMs, such as OpenAI’s GPT-4, demonstrate the ability to perform relatively sophisticated human tasks and show abilities that are sometimes difficult to distinguish from human intelligence. The aptitude to perform typically complicated human tasks quickly and effectively has naturally led to their explosive growth and implementation in many fields. Recent models have also introduced ‘multimodal’ input, allowing users to interact with the chatbot through various documents and images alongside text, further increasing their versatility. 

Prior to this new generation of multimodal LLMs, the implementation of AI into higher education has been explored \cite{zawacki2019litreview}. For instance, ‘intelligent tutoring systems’ that interact with students to learn and solve problems without a human instructor have shown promising performance \cite{schulze2000andes, steenbergen2014meta}. Current LLMs' capabilities significantly expand AI's potential applications in higher education. In physics education, recent studies have demonstrated the impact of ChatGPT on assessments and other applications, such as generating solutions to standard physics problems \cite{yeadon2023physicsessay, liang2023exploringgpt}. Despite the growing interest, the study of new LLM-based AI within physics education is still limited due to the novelty of such tools and further investigation is required. 

This study assesses AI chatbots' ability to grade and provide feedback on physics problems at the university level. The motivation for such an investigation is clear: effective grading and feedback are crucial for student learning, allowing students to gauge an understanding of their comprehension and address specific areas of improvement. Human grading and feedback is a labour-intensive process involving not only the initial grading of student solutions but also potential moderation or adjustment processes. Such a delay can reduce the speed and quality of feedback \cite{Paris_2022feedbackchallenges}. 

Automating grading with AI could substantially reduce the workload and lead to faster and more consistent feedback without compromising quality. AI could provide feedback on textbook exercises during independent study, supplementing instructor-led learning. This study aims to evaluate the feasibility and effectiveness of using LLM-based AI chatbots as an automated grading system in physics education to provide students with grading and feedback. 

\subsection{LLMs and AI Chatbots}

To thoroughly investigate how effectively AI chatbots can perform grading, it is essential to understand the core features of AI chatbots, the LLMs underlying them and their potential integration into an automated grading system. 

Polverini and Gregorcic provide an in-depth discussion on using AI chatbots and LLMs in physics education \cite{LLMinphysicseducation}. Recent LLMs, such as GPT-4, are deep learning-based neural networks trained on internet data. They process a prompt, including input text and other information types, such as images, to generate text-based responses. In text generation, a metric known as `temperature' is a degree of randomness \cite{eliot2024temperature}. As a result, inputting the same prompt twice with non-zero values of this parameter will lead to two different generated text responses. 

Since LLMs generate words based on statistical likelihoods following a prompt, they are not inherently optimised for conversational or task-specific responses. To form an AI chatbot for conversational responses and carrying out instruction-based tasks, the original prompt is altered to encourage the generation of a conversational or task-based response more likely, as discussed by Shanahan \cite{shanahan2023talkinglargelanguagemodels}. LLMs are often retrained or fine-tuned to enhance their performance in dialogue-based contexts \cite{LLMinphysicseducation}. These adaptations transform the LLMs into conversational AI chatbots capable of interactive dialogue, task execution, and instructions following a user prompt. 

AI chatbots perform relatively well within general academic assessments and intelligence metrics \cite{geminiteam2024, openai2024gpt4technicalreport, anthropic2024claude3}. However, the apparent 'intelligence' or ‘reasoning' is merely a simulated understanding, reflecting their statistical model rather than a genuine conscientious understanding of knowledge. 

Moreover, the performance of LLMs in understanding instructions and solving problems can be drastically affected by the structure of prompts \cite{chen2024unleashingpotentialpromptengineering}. Constructing prompts and instructions to improve LLM performance is known as ‘Prompt Engineering'. Since grading physics solutions will involve giving chatbot instructions through prompts and performing the task of grading and feedback, prompt engineering will play a critical role in evaluating the ability of AI chatbots to grade and provide feedback effectively. 

\begin{figure*}[t]

\includegraphics[width=1\textwidth]{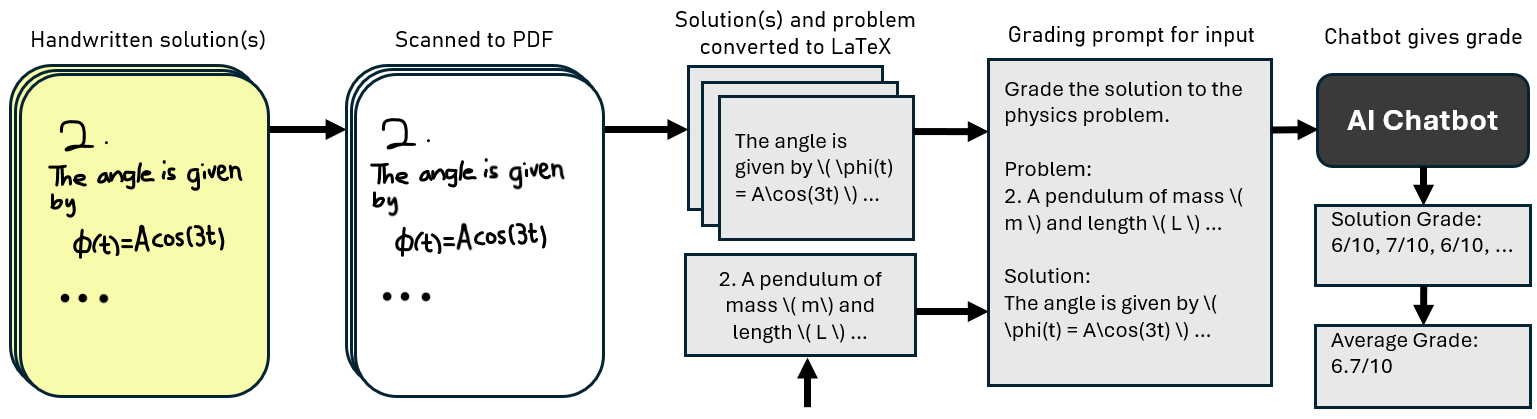}
\caption{\label{gradingsystemfigure}Diagram example of a constructed automated grading system using an AI chatbot.  }
\end{figure*}

\subsection{Automated AI Grading Systems}

 Previous research has examined the structure of automated grading systems \cite{kortemeyer1, kortemeyer2} in a typical scenario where a student submits a handwritten solution or batch of solutions for a physics problem, like in the case of exam questions. For the chatbot to accurately interpret and assess these solutions, the handwritten answers must first be digitised to ensure precise translation of both textual and mathematical content. 

Kortemeyer \cite{kortemeyer1} proposes a method: scan the handwritten solutions into a PDF format and then, using AI tools or optical recognition software, convert the PDF into a LaTeX file. Research on OpenAI's ChatGPT suggests it can effectively interpret mathematical content formatted in LaTeX \cite{frieder2023mathematicalcapabilitieschatgpt}. Thus, the handwriting and mathematical symbols can be encoded in a machine-readable language. This capability is assumed to extend to all AI chatbots. The physics problem must be provided for the chatbot to evaluate the solutions; therefore, it must be formatted in LaTeX along with a LaTeX-encoded marking scheme, which the chatbot must follow.

The LaTeX-encoded problem with handwritten solutions can then be combined into a single prompt with instructions for grading. This is then fed into the AI chatbot, which should output a corresponding grade. However, given the inherent randomness associated with the chatbot responses, it is possible that repeated input of the same grading prompt results in different grades. To address this, the grading prompt can be submitted several times, and average the answers. Such a grading system is visualised in Fig. \ref{gradingsystemfigure}.

The difference in the type of questions introduces an additional layer of complexity. Calculation-based questions may result in numerical answers without intermediate steps. In questions requiring long explanations, a student may demonstrate a sound understanding but use language and keywords that differ from the mark scheme. Such scenarios create ambiguity regarding the appropriate allocation of marks, so a thorough investigation into these subtleties is necessary to provide insight into the practicalities of using AI to automate grading.

\section{Methodology}

\subsection{General Method}

The approach to assessing grading performance involved directly comparing AI grading of physics solutions with human grading. Analysing a sufficiently large set of physics problems and their solutions makes it possible to infer statistically, or at least qualitatively, how effectively AI chatbots perform when grading physics solutions. 

The method can be chronologically split into five stages: creation of physics problems and mark scheme, generation of solutions, AI grading, human grading, and analysis of the results.  The method is implemented to evaluate the performance of four different LLMs: GPT-4 and GPT-4o by OpenAI, Claude 3.5 Sonnet by Anthropic and Gemini 1.5 Pro by Google Deepmind. All prompt inputs were submitted through the respective chatbot websites \cite{openai_chatgpt, anthropic_claude, google_gemini}.

\subsection{Recognition of handwritten text}
The first test, independent from the main body of the study, assesses the ability of ChatGPT4 to recognise handwritten language and translate it into a LaTeX document. This has been done using several handwritten old exam papers and assignments (that cannot be discosed here for privacy reasons), and presented to ChatGPT 3.5, ChatGPT4 and Gemini 1.0. Of the three models tested, only ChatGPT4 gave good results when the text converted by the LLM was evaluated by a human reader. As long as the quality of the scanning was better than 200 dpi, more than 80\% of the text was correctly converted, with the written words almost always correctly understood, but some problems in understanding formulae: sometimes extra text (like "let x be", correct but not present in the written text) was added, and often superscripts like the arrow sign indicating vectors, or the double dots indicating double differentiation with respect to time, were confused or misinterpreted. The conclusion is that handwriting recognition, while decent, is not sufficiently reliable for grading real-life exam papers. In the remainder of this study, solutions generated by the LLMs themselves will be used, composed directly in LaTeX format.

\subsection{Physics Problem and Mark Scheme Creation}

Physics problems were chosen from three main topics: Classical Mechanics (CM), Quantum Mechanics (QM) and Electromagnetic Theory (EM). As these are core areas in undergraduate physics curricula, evaluating AI grading within these three topics should provide valuable insight into the potential for implementing automated AI grading. However, it is acknowledged that large areas of physics, such as Thermodynamics and Statistical Mechanics, have been omitted, so caution is advised in generalising the results of this investigation to other topics.

A total of 10 problems in each of the three main topics were selected, resulting in a dataset of 30 undergraduate physics problems. The problems were adapted from questions in exam papers, problem sheets and coursework within relevant undergraduate physics courses at University College London (UCL). The selected problems cover a wide range of content. For instance, in the topic of EM, problems span from electrostatics to circuits to electromagnetic wave theory. Observing AI grading performance across a large span of content is intended to reveal any underlying general behaviours in grading accuracy. 

\begin{figure*}[t]

\includegraphics[width=0.8\textwidth]{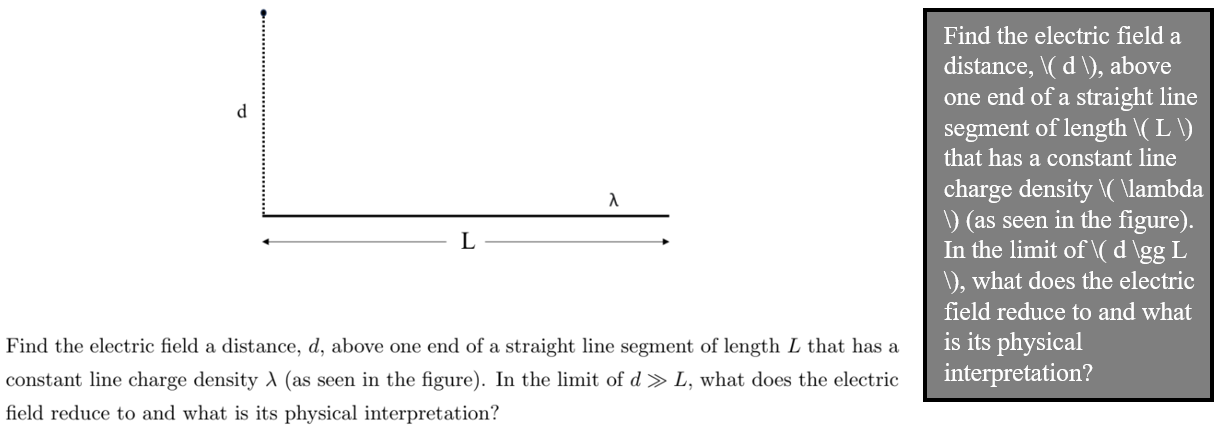}
\caption{\label{EMproblem}(Left) An example of an EM physics problem which has a corresponding figure. (Right) The same question is written in the LaTeX encoded form for prompt input. }
\end{figure*}

\begin{figure*}[t]

\includegraphics[width=0.8\textwidth]{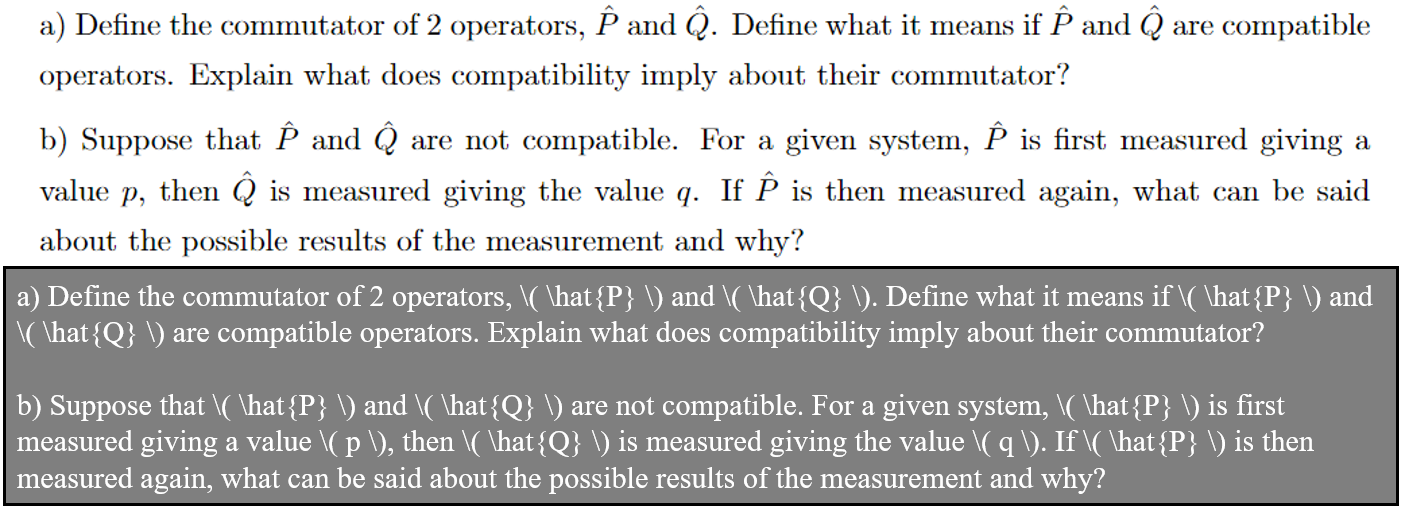}
\caption{\label{QMproblem}(Top) An example QM word-based problem which requires explanation rather than heavy mathematics to answer. (Bottom) The same question is written in the LaTeX encoded form for prompt input.}
\end{figure*}

\begin{figure*}[hbt!]

\includegraphics[width=0.85\textwidth]{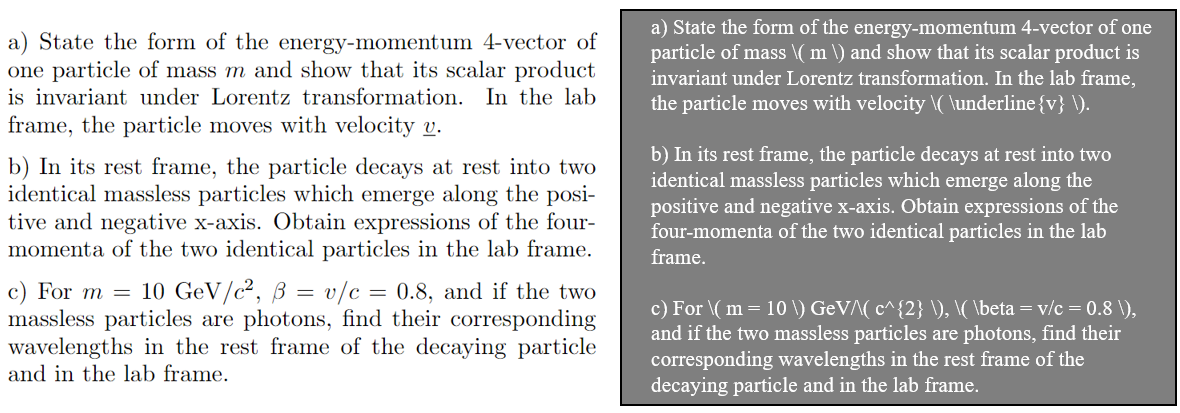}
\caption{\label{CMproblem} A standard problem on Lorentz transformations with no corresponding figure that requires mathematical calculation to solve.}
\end{figure*}

Within each topic, two of the ten problems are 'word-based', requiring a primarily worded explanation of physics rather than extensive calculation. The remaining eight problems require significant mathematics or involve calculations. This two-to-eight ratio was chosen to reflect typical undergraduate physics assessments, which include problems which require at least some level of mathematical manipulation or quantitative calculation. Additionally, six of the thirty problems, three from CM and three from EM have a corresponding figure that can be input into the AI, enabling an evaluation of multimodal capabilities and their impact on grading. Different types of problem types are shown in Fig. \ref{EMproblem}, \ref{QMproblem} and \ref{CMproblem}. 

It should be noted that many of the problems are partitioned into parts a), b), and c), such as Fig. \ref{CMproblem}. This is chosen as a prompt engineering strategy inspired by OpenAI guidelines \cite{openaipromptengineering}, enabling AI to 'view' each section of a problem clearly, potentially enhancing grading performance. 

With this chosen set of problems, an associated mark scheme is developed for every problem. The mark scheme assigns an allocated number of marks/points and one possible sample solution to each physics problem. This mark scheme is critical to the method, since it gives a framework by which humans and AI chatbots will grade. Human graders will grade physics solutions based on the mark scheme, assigning a number of marks to the solution out of the total marks possible, and the same will be done for AI chatbots. Even more importantly, the mark scheme turns grading into a numerical process, allowing quantification of AI grading performance. Other grading frameworks may be chosen, such as a rubric-based mark scheme that gives points to physics solutions based on some criteria. However, a marks/point-based grading system was chosen since this is commonly implemented within physics examinations or problem-based coursework and is more straightforward than grading from a rubric-based criteria system. 

\begin{figure*}[t]

\includegraphics[width=0.75\textwidth]{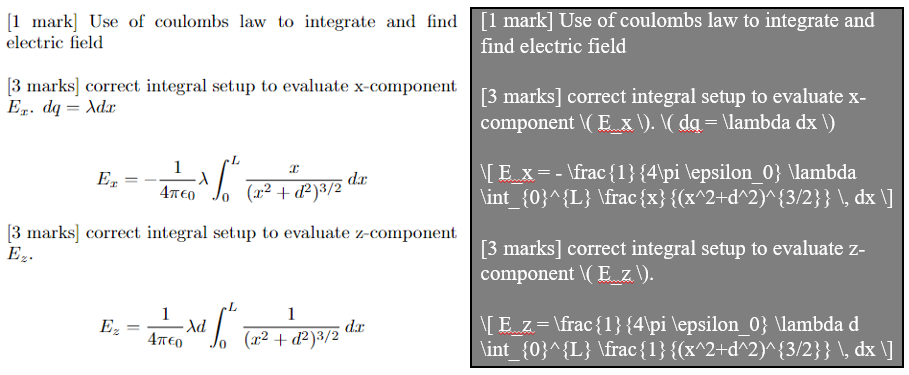}
\caption{\label{markscheme_fig} A small snippet of the mark scheme corresponding to the physics problem illustrated in Fig \ref{EMproblem}. It is allocated 14 marks in total. Its LaTeX format is also shown, which is used within prompts.}
\end{figure*}

A small sample portion of the mark scheme is shown in Fig \ref{markscheme_fig}. For questions with parts a), b), and c) style, total marks for the question are subdivided into each part. For example, for a physics problem consisting of parts a) and b) and totalling five marks, two marks may be considered part a) and three for part b). 

\subsection{Solution Generation}

\begin{figure}[b]
\includegraphics[width=0.3\textwidth]{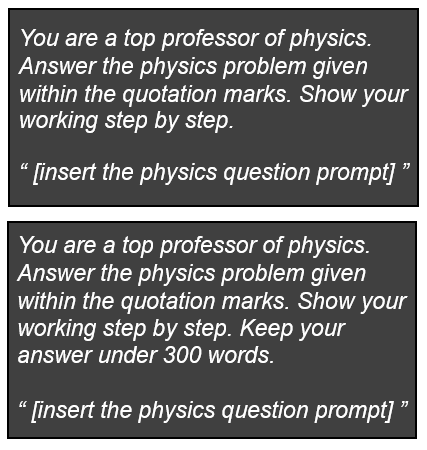}
\caption{\label{solution_prompt} Prompts used for the generation of sample solutions through ChatGPT-4. The required physics problem in the LaTeX form is inserted between quotation marks. The prompt in the top box is used for most problems in the dataset, while word explanation-type problems use the bottom prompt.}
\end{figure}

A list of physics problems and a mark scheme also require corresponding solutions to grade. Obtaining solutions from real students have ethical complications regarding obtaining consent from the students for using their work. To avoid this issue, solutions have been generated by the LLM itself,  using ChatGPT-4 (ChatGPT with the GPT-4 LLM), the model that after a preliminary study was showing the best problem-solving abilities.

A prompt is input into ChatGPT-4 through the ChatGPT website, asking it to solve a particular physics problem (the prompt form is given by Fig. \ref{solution_prompt}). The solution it generates is then copied into its LaTeX form and stored. This procedure is repeated two more times (in new separate chats) to obtain three solutions to each problem. Because of the non-zero temperature of ChatGPT-4, each solution is different from the others by varying degrees due to the randomness in response. This allows us to generate many different solutions with a relatively limited number of physics problems. This solution generation process is applied for all 30 physics problems in the dataset, giving a total of 90 solutions (3 solutions per particular  problem). For problems with figures, the figure is also inputted into the chatbot along with the original prompt. 

\begin{figure*}[t]

\includegraphics[width=0.9\textwidth]{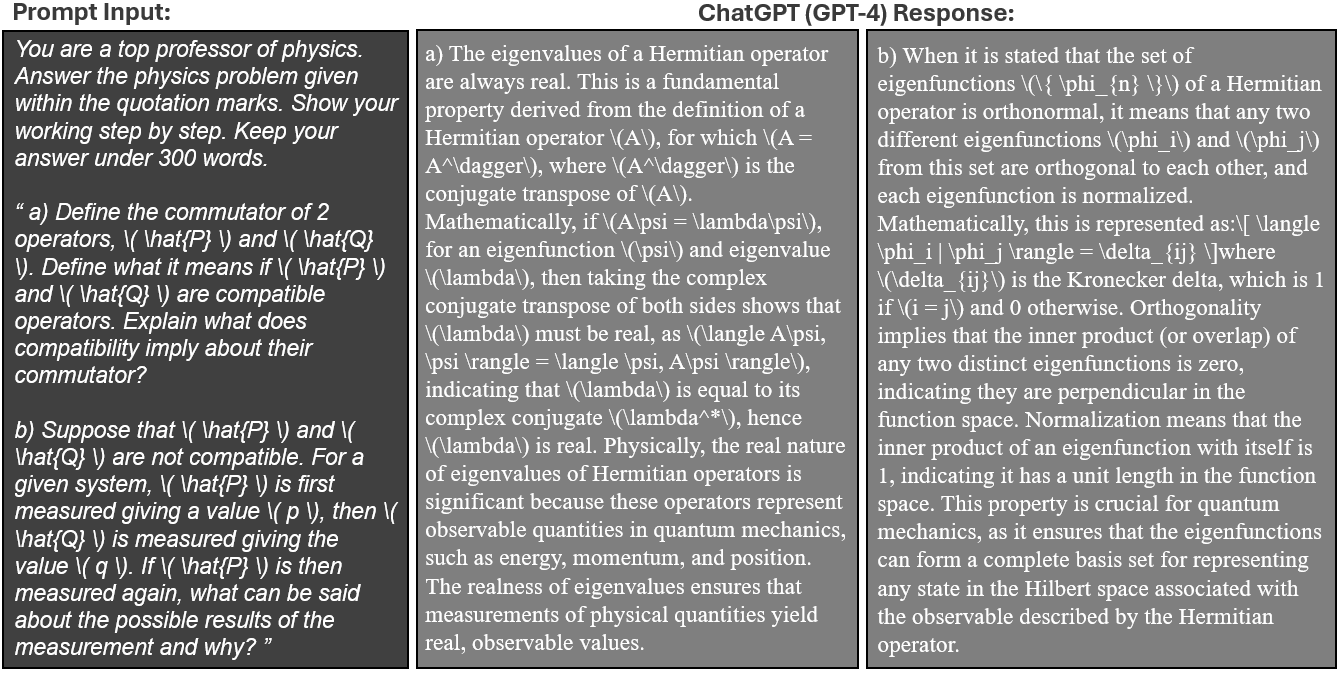}
\caption{\label{solution_example} A prompt entered into ChatGPT using the GPT-4 LLM and the solution generated by it in LaTeX form. The physics question answered is the same one outlined by Fig. \ref{QMproblem}. This solution is 1 of 90 used in grading.}
\end{figure*}

Prompt engineering plays an important role in solution generation, and the strategy implemented to generate solutions can be viewed in Fig. \ref{solution_prompt}. The approach involves persona adoption and a CoT (chain-of-thought, \cite{CoTprompting}) command within the initial prompt lines, along with an instruction to solve the corresponding physics problem. This strategy attempts to improve the response's reasoning and quality. Quotation mark delimiters are used to highlight the physics problem within the prompt to attempt to improve prompt clarity. For the 'word-based' problems, which involve heavier explanations, an additional command to keep the answer under 300 words is included to ensure conciseness. An example of a prompt used followed by the solution generated by GPT-4 is given in Fig. \ref{solution_example}. The solutions were generated between 15/02/24 to 01/03/24. 

\subsection{AI Grading of Solutions}

With a set of physics problems and a corresponding set of physics solutions, it's possible to prompt AI chatbots to grade each solution, which forms the heart of this method. Two different approaches are considered when prompting instructions for AI to grade solutions. In the first approach, the chatbot is allowed to grade solutions blindly with no mark scheme, while in the second approach, the chatbot is provided with a mark scheme to facilitate grading.

\subsubsection{‘Blind' Zero-shot Grading}

\begin{figure*}[hbt!]

\includegraphics[width=0.8\textwidth]{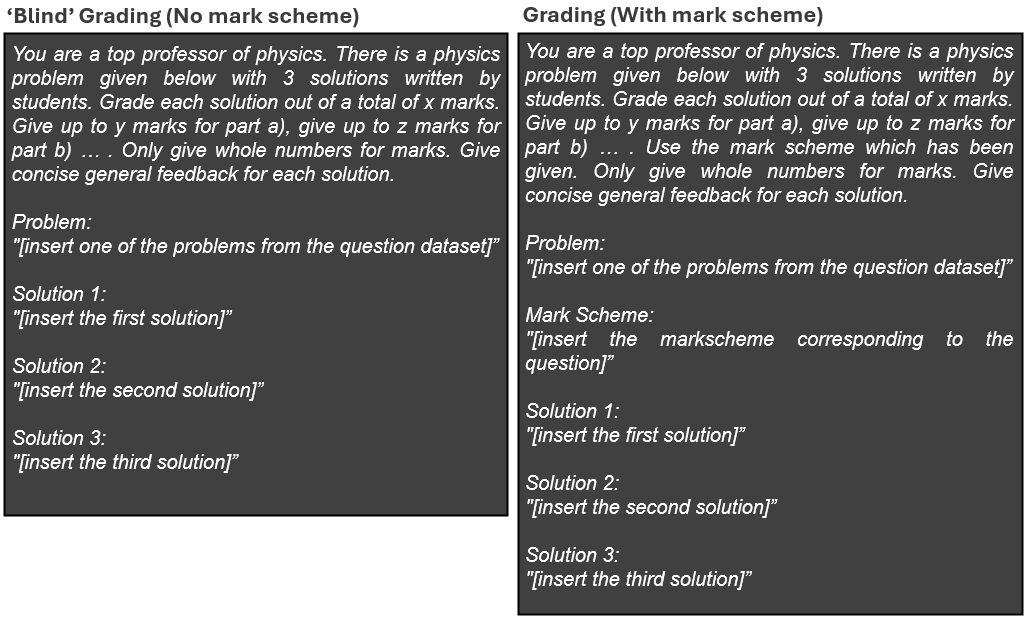}
\caption{\label{grading_prompts} The prompts used for AI chatbot grading (blind and with mark scheme). The three solutions (and mark scheme) corresponding to the physics problem are inserted into the respective quotation marks in their LaTeX form. }
\end{figure*}

\begin{figure*}[hbt!]
\includegraphics[width=1\textwidth]{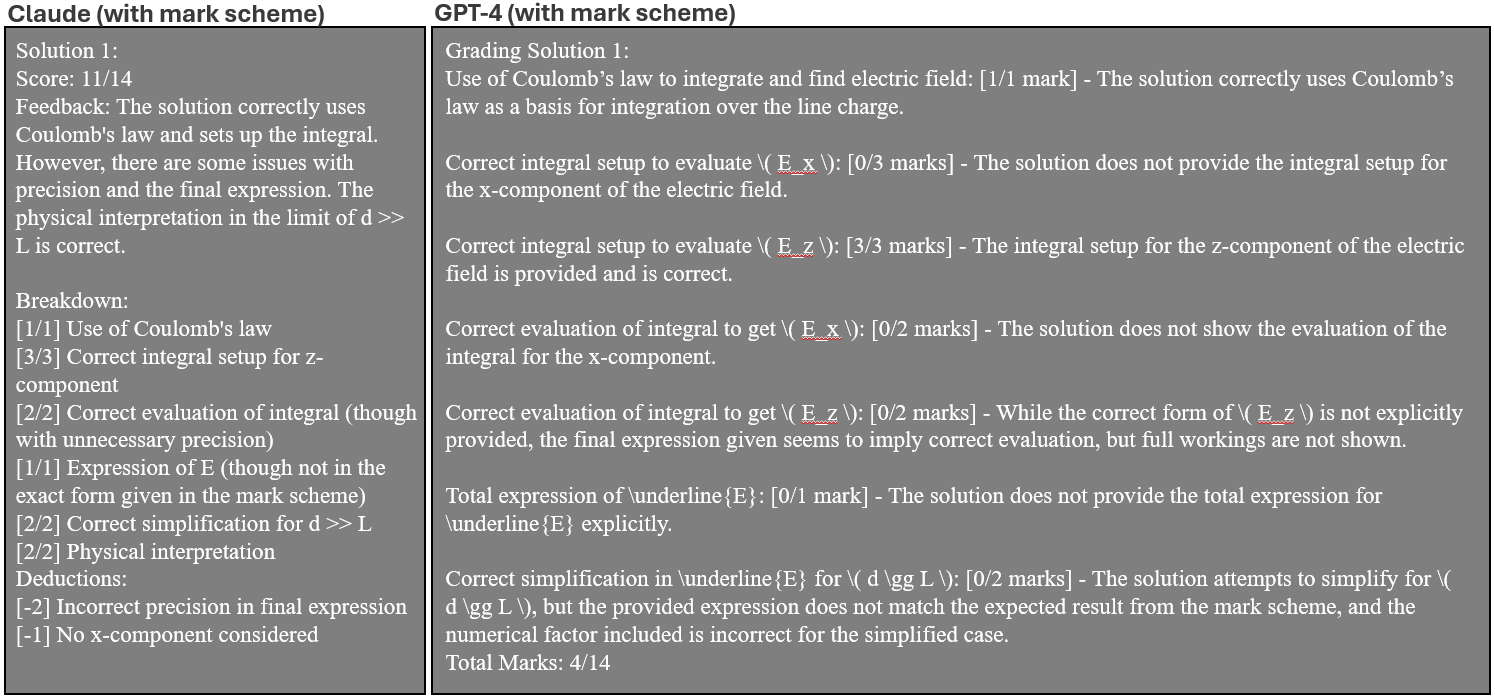}
\caption{\label{grading_examples} Two examples of the prompt responses to grade a physics problem (the problem in Fig. \ref{EMproblem}) using the mark scheme. Only a small portion of the responses are shown, such as the grading of one solution (both responses grade the same solution). There are two other solutions graded in the complete responses. On the left is the response given by Claude 3.5 Sonnet. On the right is GPT-4. One can note the difference in the styles of grading and response.}
\end{figure*}

An AI chatbot is prompted to grade the solutions corresponding to the three parts of each problem. The prompt structure is given by Fig. \ref{grading_prompts}. The grading prompt includes not only the physics problem and the three corresponding generated solutions but also the total number of marks, along with the marks allocated to each component of the problem. An instruction requiring an output of concise feedback for each solution is added. 

The mark scheme is not provided, requiring the chatbot to evaluate the quality of each solution and assign marks without additional resources, effectively ‘blindly’ grading. This results in a final grade and some general feedback on each solution. Once again, the chatbot has a degree of randomness, so the grade and feedback given to each solution may vary. The same grading prompt was input five times in total, with each trial in separate chat channels, to ensure previous results and prompts do not influence the chatbot result. The mean and standard error of the five answers is used in the final machine grading. This procedure is then repeated for all 30 questions so that all 90 solutions are graded 5 times. 

\subsubsection{Markscheme Grading}

The entire process outlined in the ‘blind grading' step is then repeated, modifying the prompt to include the mark scheme (in LaTeX form) (see Fig. \ref{grading_prompts}). Within the mark scheme quotation marks, the three additional statements were also included: 

\begin{enumerate}
    \item  Allow any equivalent expressions as long as they are reasonable (close to the same form).
    \item Allow any alternate solutions as long as the argument is correct and sufficient. 
    \item Deduct marks for inappropriate units and precision.
\end{enumerate}

This approach attempts to assert that other potentially viable solutions and algebraic expressions are equally correct and that the chatbot does not stick exactly to the mark scheme when grading, not allowing other correct methods. A snippet example of the kind of responses which followed prompt input is given in Fig. \ref{grading_examples}. Ultimately, this also produces a set of mean grades for each solution and the standard error of the mean.

The entire 'AI grading' step outlined here is applied to each of the different LLMs - GPT-4, GPT-4o, Claude 3.5 Sonnet and Gemini 1.5 Pro. In the case of GPT-4, there were only three trials of grading per problem instead of the five trials outlined above. GPT-4 grading prompting was done between 01/03/24 to 10/03/24, Claude 3.5 Sonnet between 28/06/24 to 20/07/24 and both GPT-4o and Gemini 1.5 Pro between 23/07/24 to 29/07/24. Before any analysis of results, it's important to note that in small, rare instances, the response given by a chatbot did not follow instructions properly or made some major mistakes, such as tallying marks incorrectly or repeating the prompt instead of providing an answer. In cases where marks are added incorrectly, the actual mark is used rather than the incorrectly added mark. In other cases, the prompt is re-entered in a separate chat, and the original erroneous result is not considered. In developing automated grading systems, these rare but critical errors will have to be accounted for by some checking system or improvement of the AI chatbot model itself.

In this approach, the LLM has simultaneous access to the three different solutions to the problem. The motivation for this prompt is to allow the grading of each solution relative to the others so that the scaling will be more consistent. This is akin to an examiner who has knowledge of other student solutions to a given exam problem, therefore can be consistent in the way different students are graded. Other strategies could be implemented and investigated. For instance, aside from the prompt length, one option is to input every single solution along with the corresponding set of problems into the chatbot and analyse its performance. Alternatively, only one solution can be inputted into the prompt each time. It can be presumed that this would influence grading ability and can be a step forward in continuing the work done in this paper. 

Due to their length, the LLM chatbot responses, along with the full data set of physics problems and corresponding mark scheme, cannot be included in this paper but are left as a reference\footnote{Raw data, AI responses, problems and mark scheme available at: \url{https://github.com/R-y-a-n-M-o-k/AI-Chatbot-Grading-Results}}.

\subsection{Human Grading}

A further step in the method is to have humans grade the generated physics solutions. Human subjectivity in grading is known, so moderators and second markers are common practice in examinations. To reduce the variation and bias of human grading results, have four graduate students from UCL graded all of the 90 solutions separately, assigning each solution a mark out of the total marks of the corresponding problem. Each mark given to a solution is then averaged to produce an average human grade - the average mark awarded to a solution out of the total marks of the corresponding problem. The standard error of the mean for each averaged human grade is also calculated. 

Each human grader marks the solutions based on the mark scheme developed for the physics problems. Of course, the mark scheme only gives one path to the solution, so graders would have to verify and check any different solutions or equivalent expressions. It is also up to the human graders to decide if units or stated precisions in solutions are appropriate. 

The grades assigned by human markers enable a quantitative assessment of AI grading performance. It’s important to keep in mind that human grading is not necessarily the correct way of grading - it is entirely possible that AI grading could be better. At face value, the comparison between AI and human grading only allows a relative comparison. 

\section{Results}

The average grade of each solution is converted into a percentage of the total marks for the corresponding physics problem. The resulting data is then analysed using regression and clustering methods. Firstly, the averaged AI chatbot grades for each LLM are plotted against the averaged human grades (percentages). This is done for the blind grading and mark scheme grading in separate plots (Fig. \ref{claude_plot} - \ref{gpt4_plot}). Each solution is also colour-categorised into its respective physics topic (EM, QM or CM) in case some additional clustering or pattern is visible. A linear least squares regression is calculated for each set of grading data points. x-Error bars on the horizontal axis (human grade) are not shown for clarity, but are used in the linear regression fit. 

\subsection{Regression Plots}

\begin{figure*}[hbt!]
\includegraphics[width=1\textwidth]{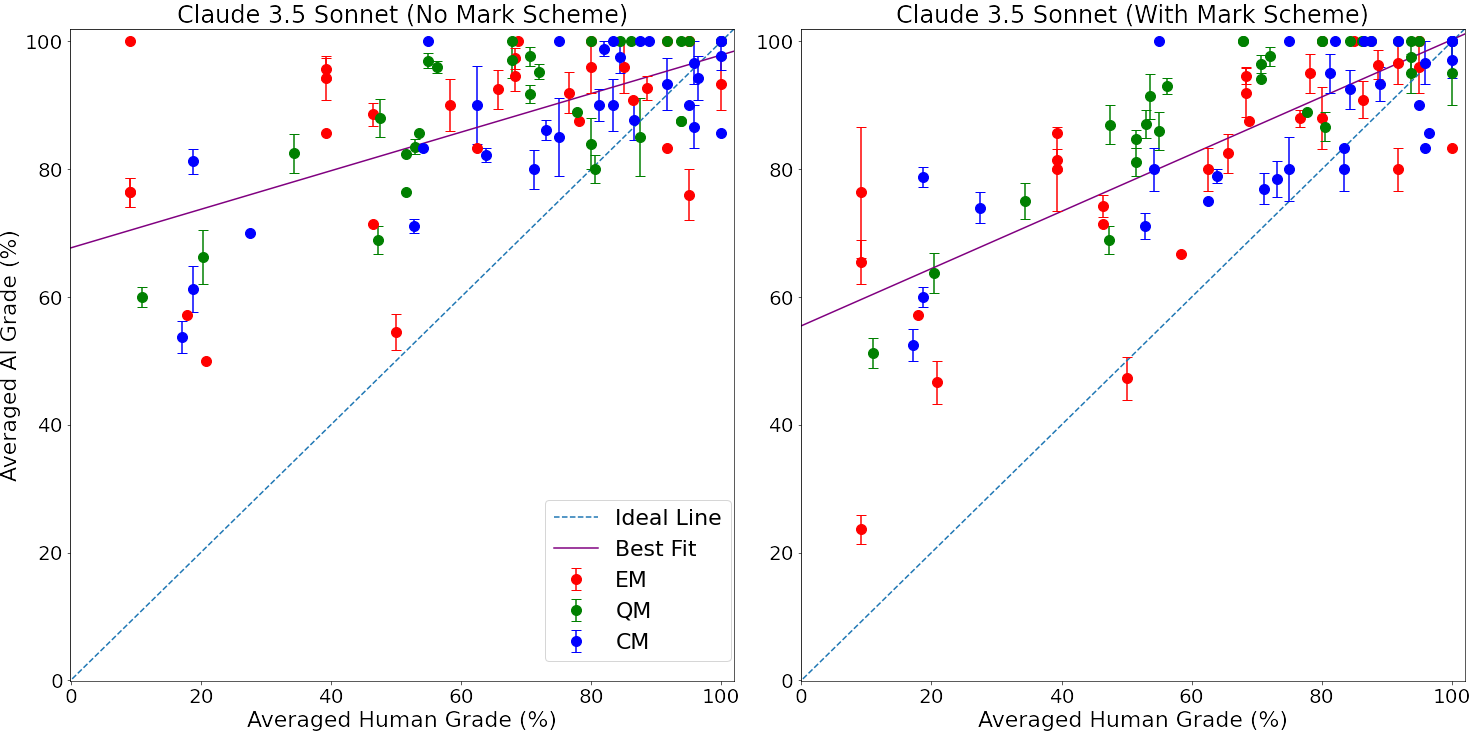}
\caption{\label{claude_plot} A plot of the average human grades against average Claude 3.5 Sonnet chatbot grades. y-errorbars are the standard error of the mean of the chatbot grade, likewise for the other plots. Left plot is grading without mark scheme, right plot is with markscheme. Datapoints may overlap exactly and, therefore, not show all 90 solutions. Without mark scheme: \(y = (0.30 \pm 0.04)x + (68 \pm 3)  \), \(r = 0.64 \). With mark scheme: \(y = (0.45 \pm 0.04)x + (55 \pm 3) \), \(r = 0.77\). Linear fits are unweighted.}
\end{figure*}

\begin{figure*}[hbt!]
\includegraphics[width=1\textwidth]{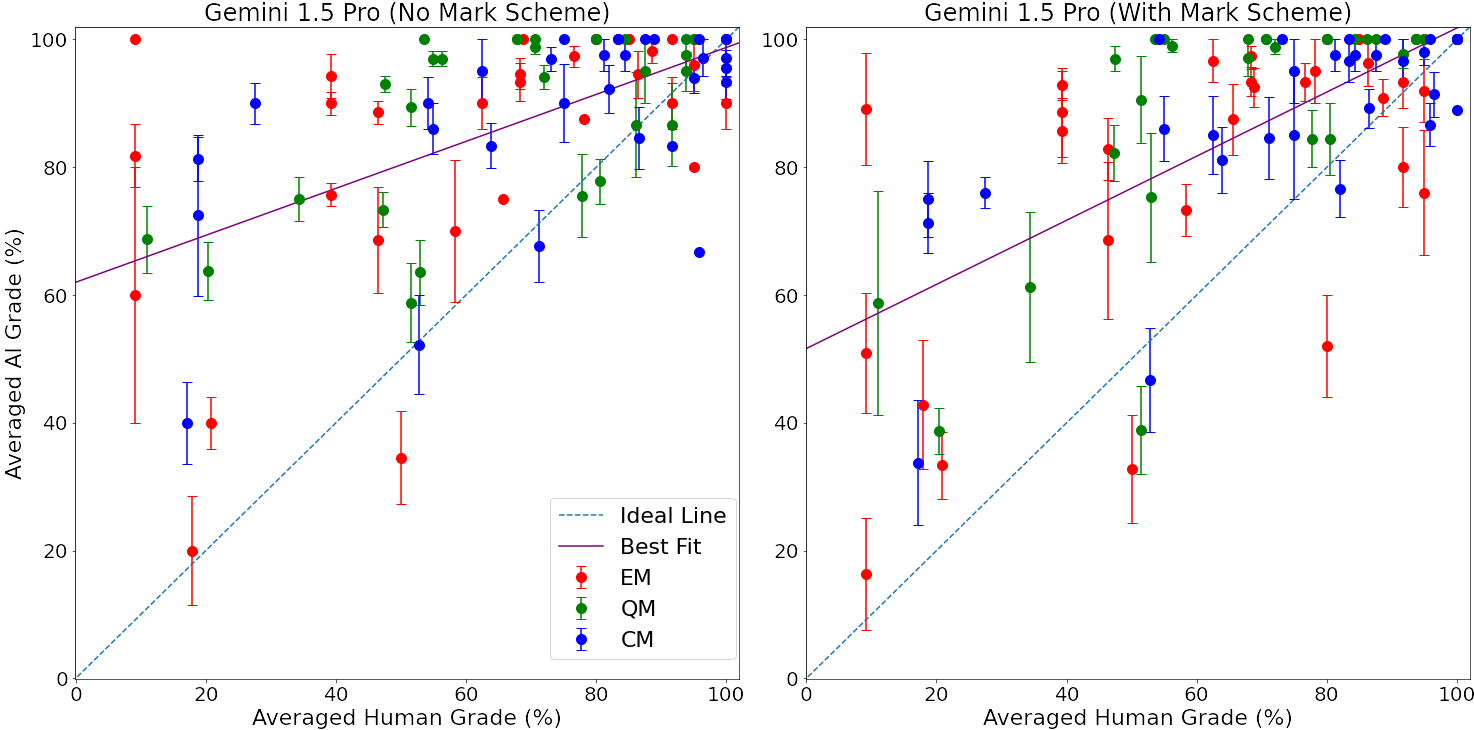}
\caption{\label{gemini_plot} A plot of the average human grades against average Gemini 1.5 Pro chatbot grades. Without mark scheme: \(y = (0.37 \pm 0.06)x + (62 \pm 4) \), \(r = 0.58 \). With mark scheme: \(y = (0.50 \pm 0.06)x + (52 \pm 4) \), \(r = 0.67\). }
\end{figure*}

\begin{figure*}[hbt!]
\includegraphics[width=1\textwidth]{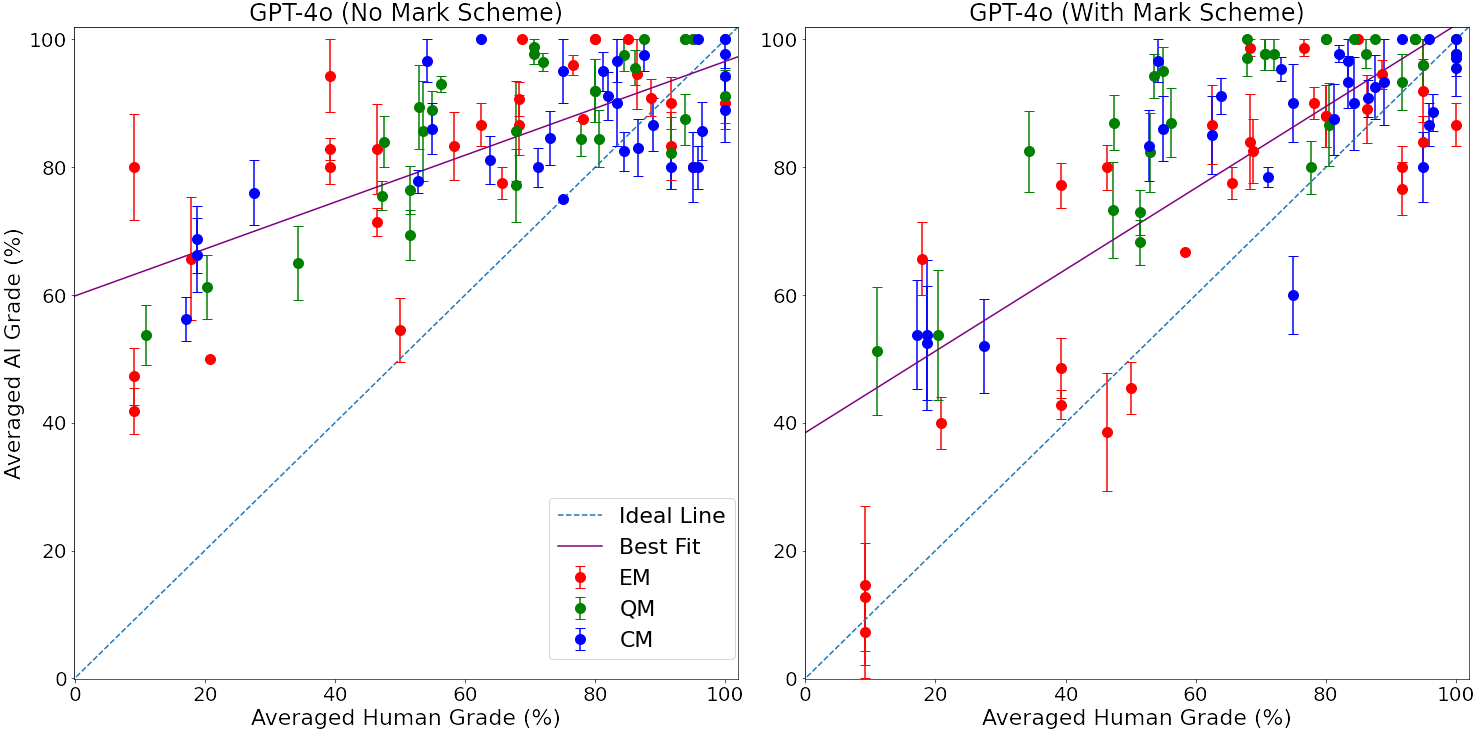}
\caption{\label{gpt4o_plot} A plot of the average human grades against average GPT-4o chatbot grades. Without mark scheme: \(y = (0.37 \pm 0.04)x + (60 \pm 3) \), \(r = 0.72 \). With mark scheme: \(y = (0.64 \pm 0.07)x + (38 \pm 4) \), \(r = 0.79\). }
\end{figure*}

\begin{figure*}[hbt!]
\includegraphics[width=1\textwidth]{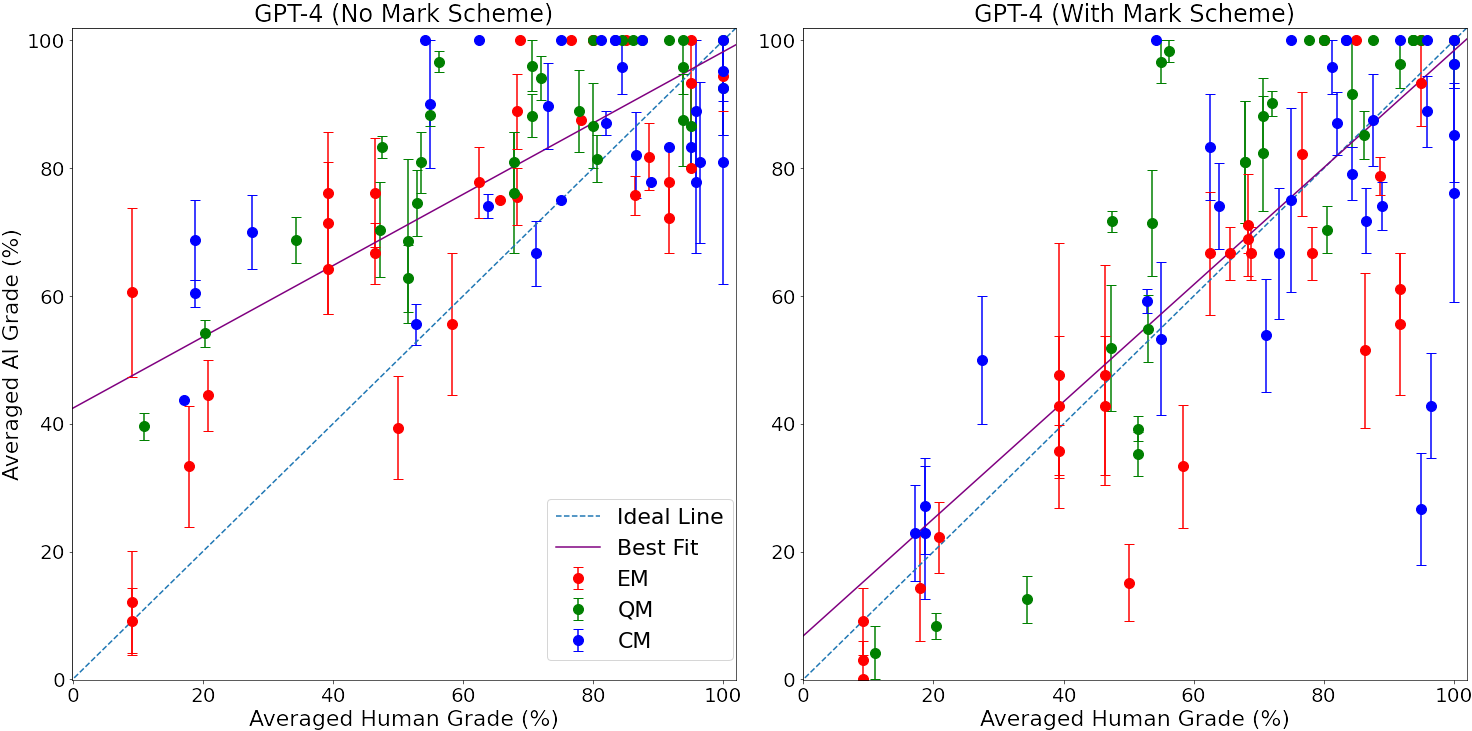}
\caption{\label{gpt4_plot} A plot of the average human grades against average GPT-4 chatbot grades. Without mark scheme: \(y = (0.56 \pm 0.05)x + (42 \pm 4) \), \(r = 0.75 \). With mark scheme: \(y = (0.92 \pm 0.07)x + (7 \pm 5) \), \(r = 0.80\). }
\end{figure*}

The plots of Fig. \ref{claude_plot}-\ref{gpt4_plot} give a simple interpretation of AI grading performance. The blue dotted line indicates the \(y=x\) function, where machine and human grades are the same. Data points above the \(y=x\) line are where solutions have been, on average, graded more leniently by the LLM chatbot relative to the human grader, while solutions below that line are graded more harshly by the AI. It should be noted the name `ideal' is somewhat fictitious since human grading is not necessarily the ideal grading, as stated before. 

\begin{figure*}[hbt!]
\includegraphics[width=0.9\textwidth]{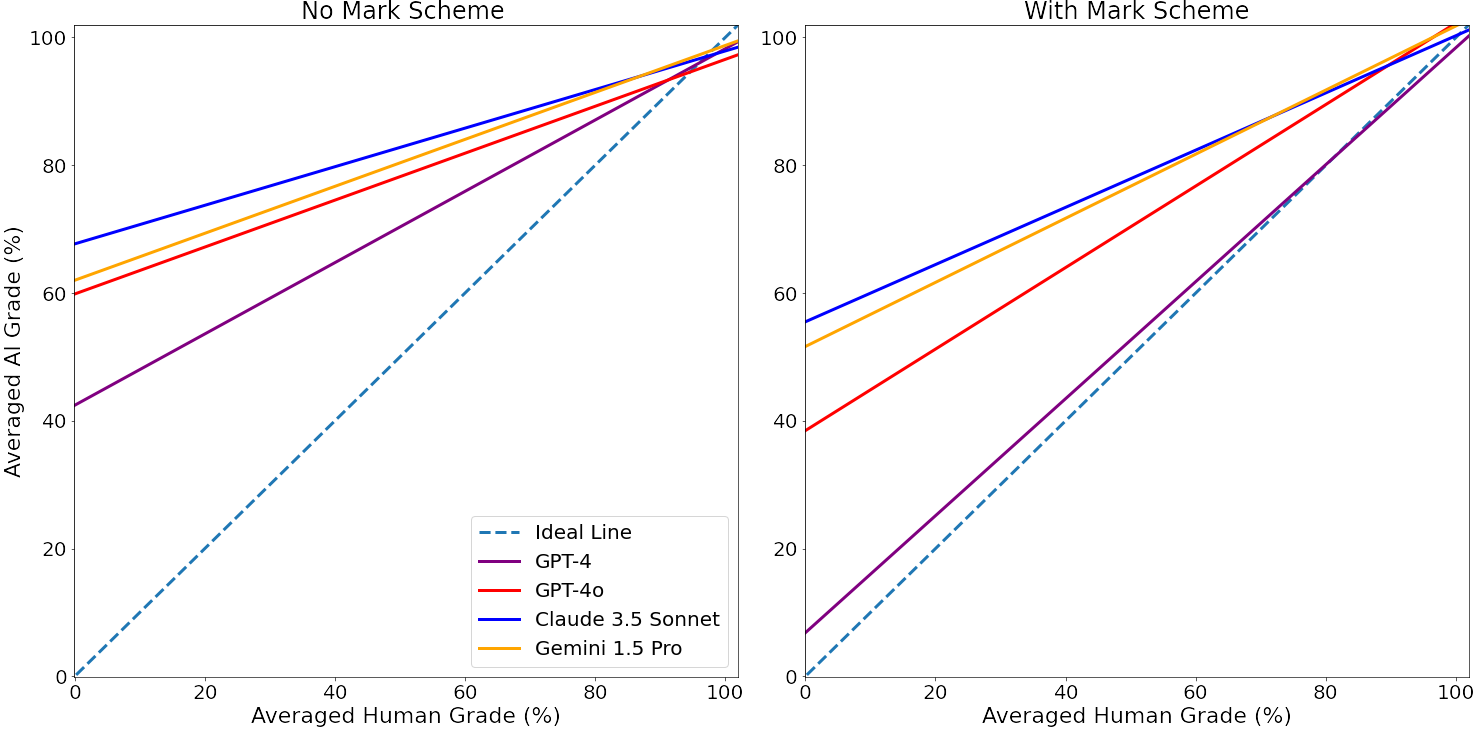}
\caption{\label{LLM_comparison} The line of best fits for each LLM plotted together.}
\end{figure*}

In general, one can see that for blind grading, all LLM models appear to be much more lenient than human graders. In order of most to least, this seems to be Claude 3.5 Sonnet,  Gemini 1.5 Pro, GPT-4o and GPT-4. The degree of leniency is more easily seen in the line of best fit and fit parameters of Fig. \ref{LLM_comparison}. Even considering uncertainty in human grading, it is fair to say that human and AI grades do not agree.

A closer  examination of chatbots' grading responses and feedback demonstrates that leniency in grading is typically a result of hallucinations, most often involving algebraic steps or mathematical expressions.

Some of the most typical examples of this behaviour are, for instance, grading to the problem seen in Fig. \ref{EMproblem}.  A correct expression for the electric field to the physics problem in Fig. \ref{EMproblem} is given by 

\begin{eqnarray}
\underline{E} = \frac{1}{4 \pi \epsilon_0} \frac{\lambda}{d} \Bigg[ \left( -1 + \frac{d}{\sqrt{d^2 + L^2}} \right) \hat{x} \nonumber \\
+ \left( \frac{L}{\sqrt{d^2 + L^2}} \right) \hat{z} \Bigg].
\label{eq:two}
\end{eqnarray}

However, in one of the generated solutions where the x-component of the electric field was left out and assumed to cancel to 0, the electric field is given as 

\begin{eqnarray}
E = \frac{0.141047395886939 \cdot L \cdot \lambda}{\sqrt{\pi} \cdot \epsilon_0 \cdot d^2 \cdot \sqrt{\frac{L^2}{d^2} + 1}}.
\label{eq:two}
\end{eqnarray}

A glance at the steps (lack of consideration for the x-contribution) and mathematical form clearly shows it is incorrect. Despite this, GPT-4 asserts 

\begin{quote}
\texttt{The final expression for the electric field is correct and the 
student appropriately considers the limit $\backslash$( d $\backslash$gg L $\backslash$), simplifying the expression to that of a point charge.}
\end{quote}

\clearpage

GPT-4o comments that 

\begin{quote}
\texttt{- Correct setup of the integral and appropriate integration steps.}
\end{quote}

with even Claude declaring 

\begin{quote}
\texttt{Accurate derivation of the electric field expression.}
\end{quote}

For the respective trials the responses are taken from, the three different models give out a total of 14 marks, 13, 11 and 13 marks, respectively, with Gemini also suggesting calculations are completely correct (without a mark scheme). There are many further instances, both more subtle or more blatant, where AI grading claims answers are correct to some extent when they are not. In some cases, while errors or mistakes in a solution are noted, the AI gives little penalty to the solution and still awards a very high mark. One could describe the attitude of AI chatbots seen in the responses as giving the benefit of the doubt to the student without any underlying reasoning. 

From the examination of the blind grading by LLM chatbots, the quality of human grading still appears substantially better, so we consider the best AI grader as the one closer to the human results. According to this criterion, GPT-4 appears to perform the best, with Claude 3.5 Sonnet performing the worst—this agrees with the qualitative analysis of the different LLM chatbot grading feedbacks. 

\subsubsection{The effect of Mark Scheme}

Introducing a mark scheme into the grading approach appears to rectify over-leniency in grading to various extents, causing the best-fit line to shift closer to the ideal line.  Gemini and Claude see some improvement with more accurate grading, however in general they are still fairly lenient relative to human grading. This is still caused by the same errors of hallucination, as can be seen in the previous Fig. \ref{grading_examples}, with the response by Claude to the same example physics problem (Fig. \ref{EMproblem}.) and the same physics solution previously discussed. Claude gives some of the statements:

\begin{quote}
\texttt{[1/1] Expression of E (though not in the exact form given in the mark scheme). \newline [2/2] Correct evaluation of integral (though with unnecessary precision) \newline [-2] Incorrect precision in final expression. \newline [-1] No x-component considered.}
\end{quote}

Once again, the correct expression for the electric field (Eq. 1) is clearly not the same as the one given by the solution (Eq. 2). Despite this, Claude asserts the expression of \(E\) is correct, even though the form is not an equivalent expression. Moreover, Claude gives very high marks for correct evaluations of integrals even though the x-component is wholly neglected,  leading to a total of 11/14 marks, which is very high considering a major error in the solution - assuming the x-component of the electric field cancels to zero. 

The introduction of the mark scheme to GPT-4o sees a slightly more sizeable effect, shifting the best-fit line by a larger amount. Thie use of the mark scheme on GPT-4 causes by far the largest shift to the line of best fit, placing it very close to the \(y=x\) line. In fact, the use of the mark scheme for GPT-4 shows a new behaviour, with many cases when the LLM is marking harsher than the humans. An analysis of these case shows that this is mostly caused by GPT-4 failing to recognise equivalent solutions or sticking to the exact solution given by the mark scheme too rigidly. An example of this behaviour is visible in the solution and grading of a classical mechanics problem: a mass (\(m\)) attached and hanging from a rope which is wrapped around a circular disk (of mass \(M\)) which can rotate, and to calculate the acceleration (\(a\)) of the mass, tension (\(T\)) and angular acceleration (\(\alpha\)). The correct expressions for \(a\), \(T\) and \(\alpha\) are given by Eq. 3. 

\begin{eqnarray}
a = \frac{mg}{m + \frac{1}{2}M} \nonumber \\
T = \frac{1}{2}Ma = \frac{\frac{1}{2}Mmg}{m + \frac{1}{2}M} \nonumber \\
\alpha = a / R = \frac{mg}{(m + \frac{1}{2}M)R}
\label{eq:three}
\end{eqnarray}

One of the solutions correctly derives the desired expressions and arguably deserves full or close to full marks. However, they were written with an equivalent expression:

\begin{eqnarray}
a = \frac{2 \cdot \frac{mgM}{M + 2m}}{M} \nonumber \\
T = \frac{mgM}{M + 2m} \nonumber \\
\alpha = \frac{2 \cdot \frac{mgM}{M + 2m}}{M R}.
\label{eq:four}
\end{eqnarray}

Although the expressions may not necessarily be the most simple, one can quickly verify that expressions for \(a\), \(T\) and \(\alpha\) are the same as Eq. 3. Despite this, GPT-4 gives the grading:

\begin{quote}
\texttt{Correct relation between linear and angular acceleration: 0 marks (this was incorrectly derived) \newline Solving simultaneous equations: 0 marks (the final expressions are incorrect)}.
\end{quote}

GPT-4 was not capable of recognising an alternate but equivalent expression of the answer and a solution path that was slightly different from the mark scheme, and only 2/7 marks were awarded.

In general, deviations and disagreements between AI and human grading are usually caused by a lack of recognition for mathematics, both in instances where solutions are correct but in different equivalent forms or when solutions give the wrong mathematical expressions. Even considering AI and human error bars, grades assigned to solutions to the mathematical part of the problem often disagree, aparty from GPT-4, which does substantially better than the other LLM chatbot models. An analysis of the plots by topic or problem type does not reveal any real underlying patterns or classifications.

\begin{figure*}[hbt!]
\includegraphics[width=0.9\textwidth]{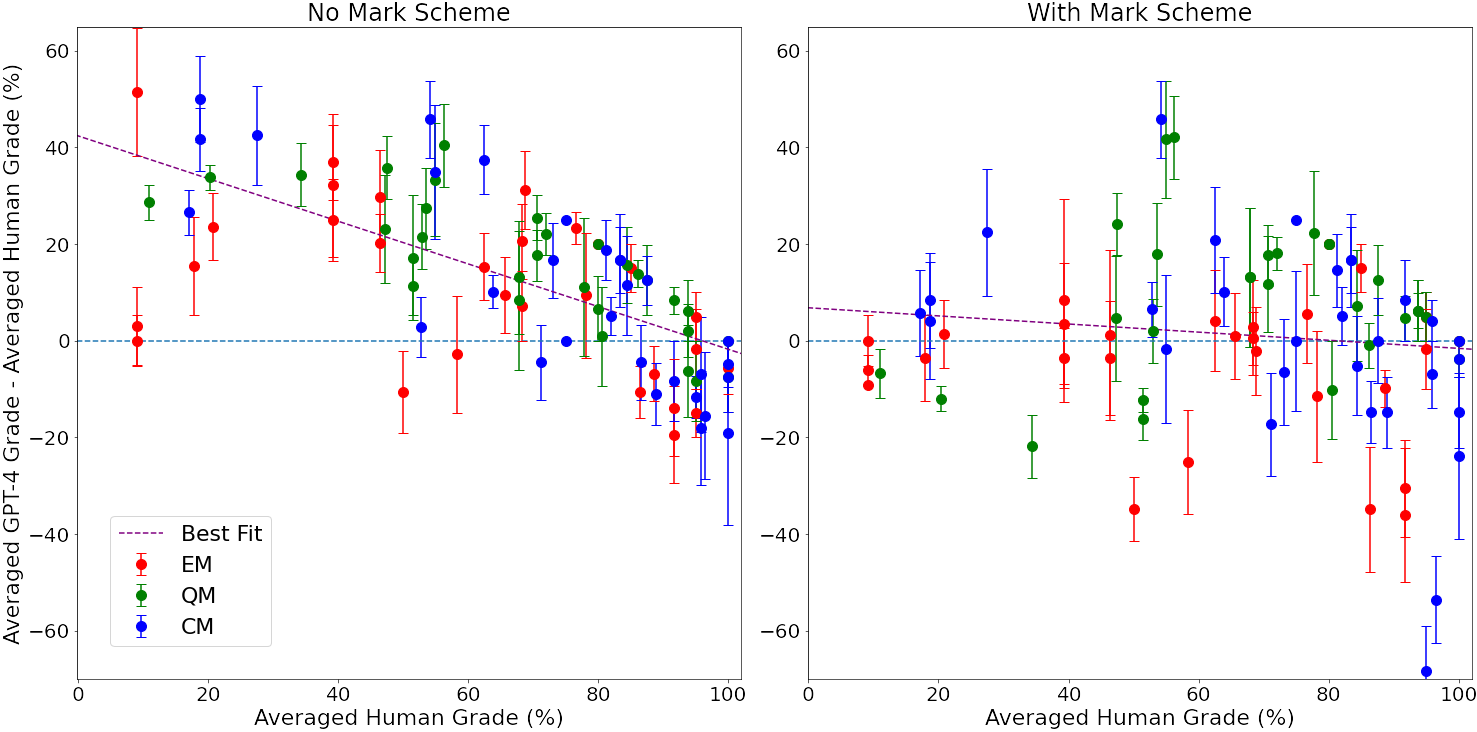}
\caption{\label{gpt4_deviation} A plot of the difference between averaged GPT-4 and human grade against averaged human grade (GPT-4 grading with and without the mark scheme). Without mark scheme: \(y = (-0.44 \pm 0.05)x + (42 \pm 4) \), \(r = -0.66\). With mark scheme: \( y = (-0.08 \pm 0.07)x + (7 \pm 5) \), \(r = -0.12\).}
\end{figure*}

\subsubsection{Grading Consistency}
The various models vary in the consistency, as indicated by the vertical error bars (reflecting the variation of grades given to a solution by a particular LLM chatbot). For example, Claude shows little or zero deviation in the grade assigned to a solution in each trial, while, for instance, GPT-4o error bars are much larger, showing bigger variation in the grade assigned. This reflects the temperature settings of each chatbot rather than a measure of consistency in grading; therefore, error bars for AI grading only retain meaning as a statistical measure of where the mean grading value lies. Assuming human grading is `ideal', a better measure of grading consistency is through the Pearson correlation coefficient, \(r\). The correlation coefficient indicates the spread of data points about the best fit. Therefore, for two different solutions which possess the same averaged human grade, \(r\) gives a general measurement of how different the AI grade for these two solutions can be - a measurement of grading consistency between different solutions and physics problems. 

Gemini shows the biggest spread of data points and the least consistency in grading with \(r = 0.58, 0.67\) for blind and mark scheme grading, respectively. With no mark scheme, Claude is also relatively inconsistent with \(r = 0.64\). The other models are generally within a range of r = \(0.7\) to  \(0.8\), namely GPT-4, with a mark scheme having the most grading consistency at r = \(0.8\). It can be observed that with the use of the mark scheme, the correlation coefficient increases, and LLM grading becomes slightly more consistent and less distributed. This is not surprising since grading based on a mark scheme is a more systematic approach. On average (per LLM), the increase in r-value is 0.085.

\subsection{AI Feedback}
A trend that can be noticed in the responses which differ between each LLM model is the use of mathematical expressions. The better-performing LLMs (GPT-4 and GPT-4o) regularly use and comment more about mathematical expressions, including larger equations or algebra, while Claude and Gemini are typically limited to small algebraic expressions such as variables, \(\alpha\), \(\phi\) or \(x\). Generally, feedback to solutions includes fairly generic statements such as ``the solution is very well written, and the calculations are correct". As expected, feedback is often skewed by the outcome of the marks and errors pointed out in the solution itself. For example, many times a solution to a physics problem which is given full marks would have feedback commenting how the answer is good even though in reality it may not be.  

An interesting example of feedback is that by Gemini on the previously discussed EM problem example. It gives statements such as 

\begin{quote}
\small{
\texttt{1. The numerical value for the electric field could be simplified further to \( \frac{\lambda}{2\pi\epsilon_0 d} \) \newline 2. The physical interpretation provided is accurate, but it could be expanded to mention the concept of a multipole expansion where the first term represents the monopole contribution \newline 3. The only minor improvement could be to explicitly state the assumption of \( dl = dx \), which holds for the given geometry where the line segment lies along the x-axis.}
}
\end{quote}

The first statement is incorrect in that the coefficient should be 2 when it should be 4 (see Eq. 1), but the idea of expressing algebraic terms rather than messy numerical values (Eq. 2) is a valid point. The second statement is effectively irrelevant especially in the context of grading an exam solution, however for a student doing an exercise attempting to improve their understanding, the mention of a multipole expansion is a different perspective which a student may not have thought of and find useful. The third statement, although subtle, is a good point of habit if \(dl = dx\) is not explicitly stated. From this example, one can see that technical errors can limit the effectiveness of AI feedback but, at times, demonstrate contextually good advice and approaches that a student may find helpful. 

Although a student will not be able to learn new content from a chatbot due to the frequency of errors made, a strong student who may already have enough knowledge to distinguish between correct and incorrect information may be able to identify different perspectives given by an AI that they may not have previously considered and, in this sense, find a chatbot to be a useful tool.

\subsection{Grade Rescaling}

\begin{figure*}[hbt!]
\includegraphics[width=0.9\textwidth]{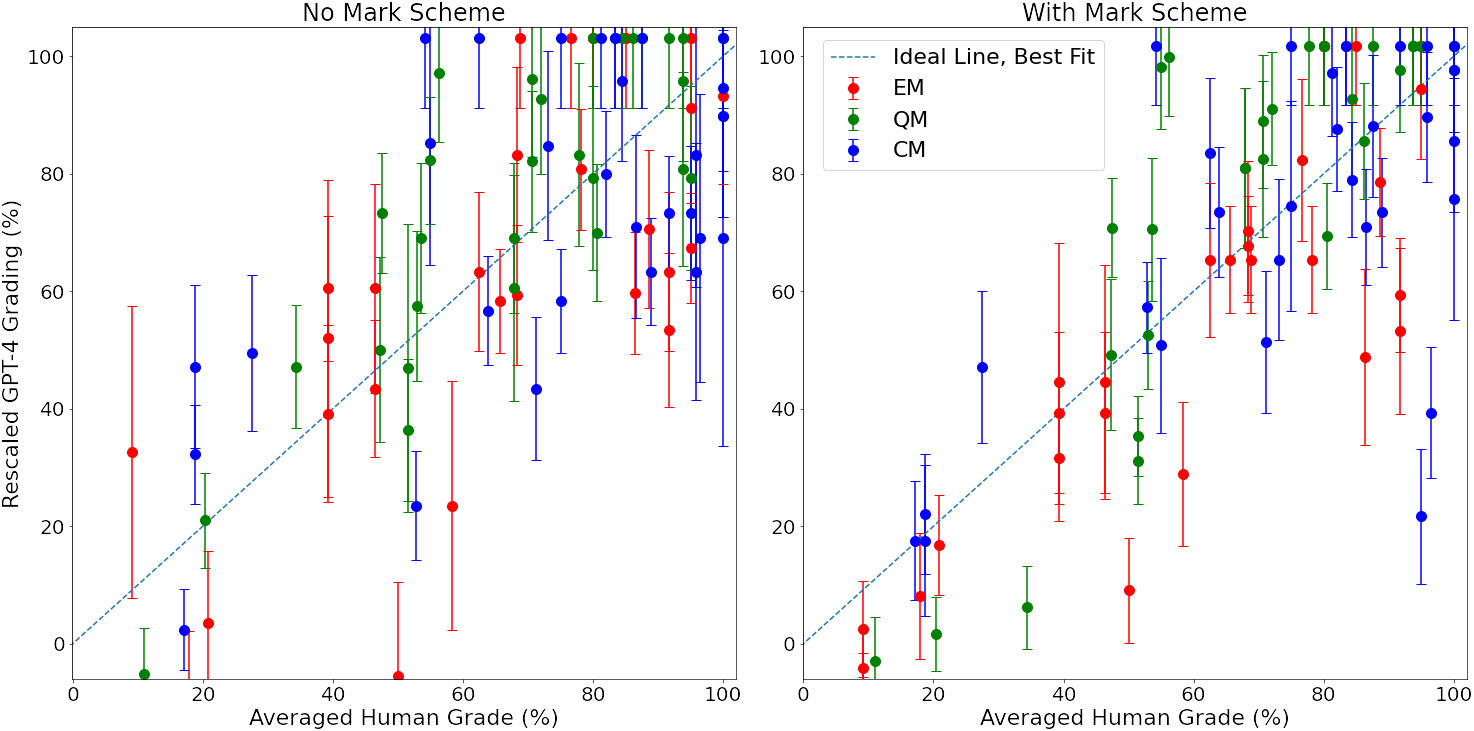}
\caption{\label{rescaled_grades}Rescaled GPT-4 grading for blind and mark scheme grades. For given original line of best fit \(y' = mx+c\), the y-datapoint is given by \(y = (y'-c)/m\). The line of best fits is the \(y=x\) line. Notice that data points are still spread out and that certain rescaled grades are above 100 or below 0. The y-uncertainties are propagated, and x-errorbars would be the same as Fig. \ref{gpt4_with_x}. }
\end{figure*}

To replicate human grading, under the assumption that human grades are accurate for each solution, each AI grade can be rescaled by a fixed amount. For a given LLM, by subtracting the AI grade by the y-intercept of the line of best fit and multiplying by the reciprocal of the gradient, the `rescaled' grades would by construction have the line of best fit \(y=x\), and are an improved AI prediction of grading values. This process is, therefore, a cheap correction one may perform to get a more accurate agreement with human grading and could be a process worth implementing in a future AI chatbot-based automated grading system. 

However, the grade rescaling does not correct for the original dispersion and grading inconsistency. Even though corrected AI grades follow a \(y=x\) line of best fit, they can have a small r-value and a large spread of data points. This approach is of practical use only if the LLM solution shows high r-values before rescaling, and the spread in grading data points agrees with human grading uncertainties. Furthermore, feedback is not modified by the rescaling, and can become incompatible with the corrected grading. Rescaling can also cause full-mark solutions to be above 100\% or zero-mark solutions to be less than 0\%. An example of rescaling AI grading is performed on GPT-4 and is shown in Fig. \ref{rescaled_grades}, where we notice that, despite both plots have on average Ai grades closer to the human ones, the left plot, indicating marking done without the marking scheme, that was rescaled much more, has a much larger spread, and many extreme rescaled grades, and the situation is still worse for other LLM models. We conclude that in the current status of development, rescaling is not a very good solution to have more reliable machine grading performance.

\subsection{Connection between AI Problem Solving and Grading}

What is the underlying cause of AI grading poorly? When asking this question for human graders, it could be argued that if they do not understand the physics problem and lack resources for grading (such as a mark scheme), they are less likely to provide the correct mark for a given solution. This motivation leads to an exploration of the same question for AI.

Specifically for the case of GPT-4, the x-axis represents the averaged human grade, while the y-axis shows the difference between the average GPT-4 (chatbot) grade and the average human grade (Fig. \ref{gpt4_deviation}). Only for GPT-4, the human grading for each solution has a second meaning - not only does it show what humans grade each physics solution, but if one assumes the human grades are an accurate representation, it also quantifies how well GPT-4 has solved each corresponding physics problem. The y-axis under the same assumption also indicates how well GPT-4 has graded each solution. 

One can see from Fig. \ref{gpt4_deviation} that there is a correlation between GPT-4 ability to solve a problem and its ability to grade. At low values of average human grade where GPT-4 has solved the problem poorly, the deviation between human and GPT-4 grades is high, and GPT-4 grades poorly. When an average human grade is higher - where GPT-4 has solved the problem well - the difference between human and GPT-4 is relatively much lower, indicating that GPT-4 is grading more accurately. Then, upon applying the mark scheme, this behaviour is effectively removed, as indicated by the near flat line of best fit. 

The behaviour of the graphs can be explained through the idea that GPT-4's ability to grade a problem is caused by how well GPT-4 can solve that particular problem. When GPT-4 cannot, or struggles, to solve a physics problem, then it should not be able to blindly grade any solutions to the problem effectively, and the difference between the grade it assigns and the grade humans assign should be large. If GPT-4 solves the problem well, it should know which solutions are good or not, and the grade it assigns should be more comparable to human grading. If GPT-4 is assigned a mark scheme, this dependence on intrinsic problem-solving ability is removed since it can rely on grading against the mark scheme if it cannot gauge the solution provided. As a result, the difference between human and GPT-4 grading can be expected to remain similar, regardless of whether GPT-4 can solve the problem well.

The data gives some evidence that there is a causal connection between an LLM's ability to solve a problem or understand how to solve a problem and its ability to grade. The argument provided is not completely definitive. If blind grading GPT-4 were to mark all solutions as good regardless of how well it can actually solve them, you could also reproduce the behaviour of the left plot of Fig. \ref{gpt4_deviation}. There are ways one can eliminate this possibility.  By generating solutions from other LLM models (e.g. Gemini, Claude, etc.), then for a solution which GPT-4 solves well (high human grade), ideally, the other LLM solution(s) should solve the problem poorly (low human grade). Then the same plot can be made and if GPT-4 manages to also grade other LLM solutions as poorly done, the difference between GPT-4 and human grade is small, this would eliminate the idea GPT-4 is just grading leniently. If GPT-4 was awarding full or near-full marks for all solutions, the poor solution would be graded highly, and the difference between human and GPT-4 grading would be larger. This can be done as an extension to the method in this paper and can also be applied to other LLM chatbot models. 

The discussion refrains from claiming a connection between `understanding' physics and grading ability for physics solutions since problem-solving ability and understanding physics are not necessarily the same thing.

\subsection{Clustering}

\begin{figure}[hbt!]
\includegraphics[width=0.5\textwidth]{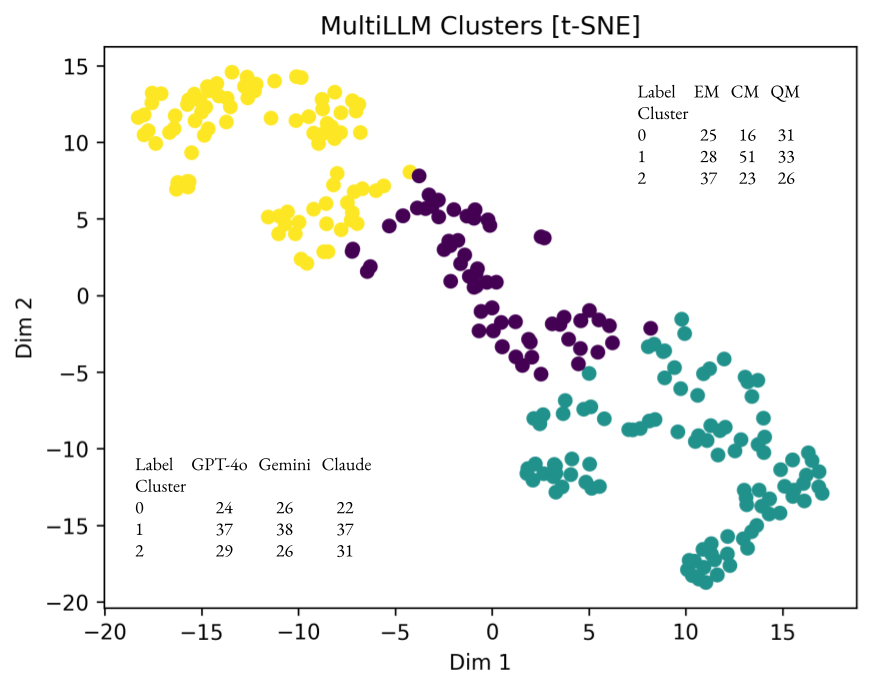}
\caption{\label{cluster} A clustered plot of t-SNE reduced data for GPT-4, Claude and Gemini. Via elbow plot, K = 3, and it is found that classical mechanics solutions are usually clustered in the bottom right cluster (cluster 1).}
\end{figure}

There is reason to believe that if LLMs assign grades to answers similarly, this would be highlighted using an unsupervised learning method such as k-means clustering. 

To ensure minimal loss of information, a table consisting of normalised grades for each grading type, a normalised standard deviation for each grading type, and the difference between the mean of the two AI grading types and the human grading is taken. This table is reduced to two dimensions using t-SNE and clustered using k-means clustering. An elbow plot estimates how many clusters the reduced data should have.\footnote{Code, analysis, clustering available at: \url{https://github.com/faraazakhtar185/LLM_Grader_Analysis}}

When all LLMs are considered together, questions pertaining to classical mechanics seem to cluster in one place. This is shown in Fig. \ref{cluster}. To understand why this may be the case, further studying the differences between the answers to classical mechanics questions and other sub-disciplines would be useful. In this case, unsupervised learning algorithms may also come in handy. 

It’s important to understand that t-SNE does not preserve global structure. Therefore, the cluster number does not indicate that a particular cluster is graded higher than another.

\section{Conclusion}

A procedure has been developed to quantify and describe the behaviour of how large language model-based chatbots may grade solutions to physics problems. The results suggest that AI grading of solutions to undergraduate-level physics problems is currently not feasible without any additional resources. Too many mathematical errors associated with hallucinations reduce the quality of and consistency substantially relative to human grading. This typically results in an unreasonable amount of over-leniency in grading. With an implementation of a mark scheme to assist LLM AI grading, although much of the grading still disagrees with human grading, the results show substantial improvements in accuracy and consistency of grading - in particular, GPT-4. For future developments of an LLM chatbot automated grading system, this seems like a promising approach for such implementation. For instance, OpenAI's new o1 appears as a potentially new candidate for investigation. 

Short term, the results are useful in gauging the effectiveness and performance capability of current LLM models in grading physics. In the long term, with how rapidly AI advances and better-improved LLMs constantly being introduced, the method established has become more and more valuable for testing new LLM-based chatbot models. The results of this investigation also suggest a link between LLM ability to solve problems n physics and its ability to grade physics solutions well. In deciding what kind of problems can be feasibly graded by AI chatbots, one may consider how well the AI can solve the problem. To improve AI's performance in grading physics problems, enhancing the problem-solving, `understanding', and mathematical ability of LLMs can be the next step in enabling much better grading capability. The methods utilised here can also investigate LLM grading ability in other fields of study, such as mathematics or chemistry.

There are many steps of the procedure that can be refined or can be explored with further investigation. First the topics in physics are limited to just three, one would also like to inspect whether grading performance for topics such as thermodynamics show similar or different behaviour. These topics are also important within an undergraduate physics curriculum, and other fields, such as chemistry or engineering, therefore, should be explored. It can also be noted that the process of collecting data through the website interfaces (such as the ChatGPT website) is fairly tedious, and this would be much easier through direct access to the API, but it was not done due to external circumstances. The use of APIs would allow investigation of how temperature settings may affect grading ability, which was not done in this paper, only using default website values.

By far, the most unknown and difficult-to-control aspects of the method are the prompting approaches and prompt engineering in general. One can expect that different strategies, such as inputting all 30 problems and 90 solutions at once, would show different grading performance than just three solutions for one problem at a time. The performance of entering multiple prompts in a single chat can also be considered. Moreover, it is very difficult to know if the performance of an LLM has been optimised or if it can perform any better. A specific aspect worth exploring would be the investigation of the performance of a few shot prompting, namely as part of the grading prompt, including several examples of grading done by a reliable human grader. One can imagine inserting examples into prompts where severe errors in solutions are recognised and harshly marked which can improve AI performance, however this will not override mathematical hallucinations of an LLM if they are not specifically pointed out in examples, therefore this is still a limitation which must be intrinsically fixed.


\nocite{*}

\bibliography{manuscript}

\appendix

\section*{Appendix}

\begin{figure*}[hbt!]
\includegraphics[width=1\textwidth]{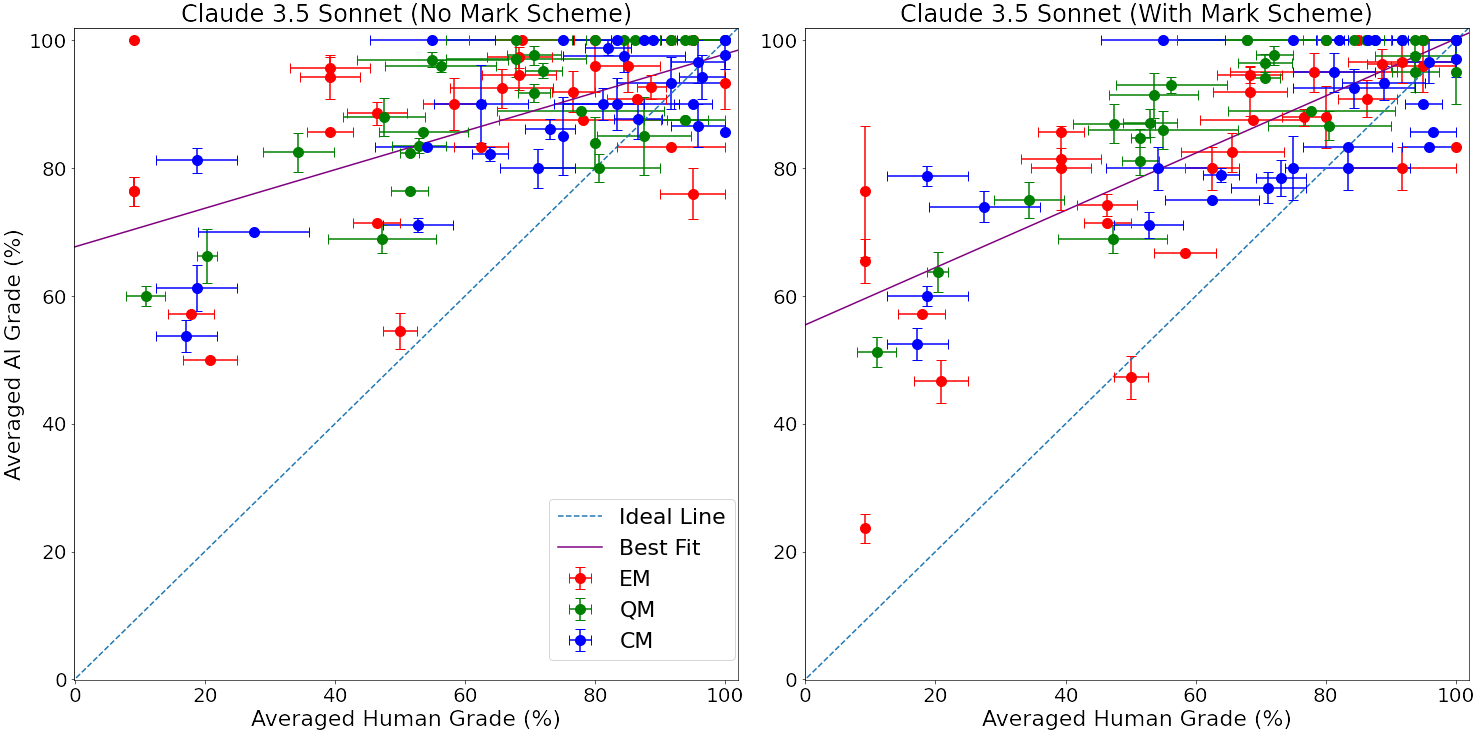}
\caption{\label{claude_plot_withx} Regression plot of Claude 3.5 Sonnet including x-errorbars. (Same as Fig. \ref{claude_plot}).}
\end{figure*}

\begin{figure*}[hbt!]
\includegraphics[width=1\textwidth]{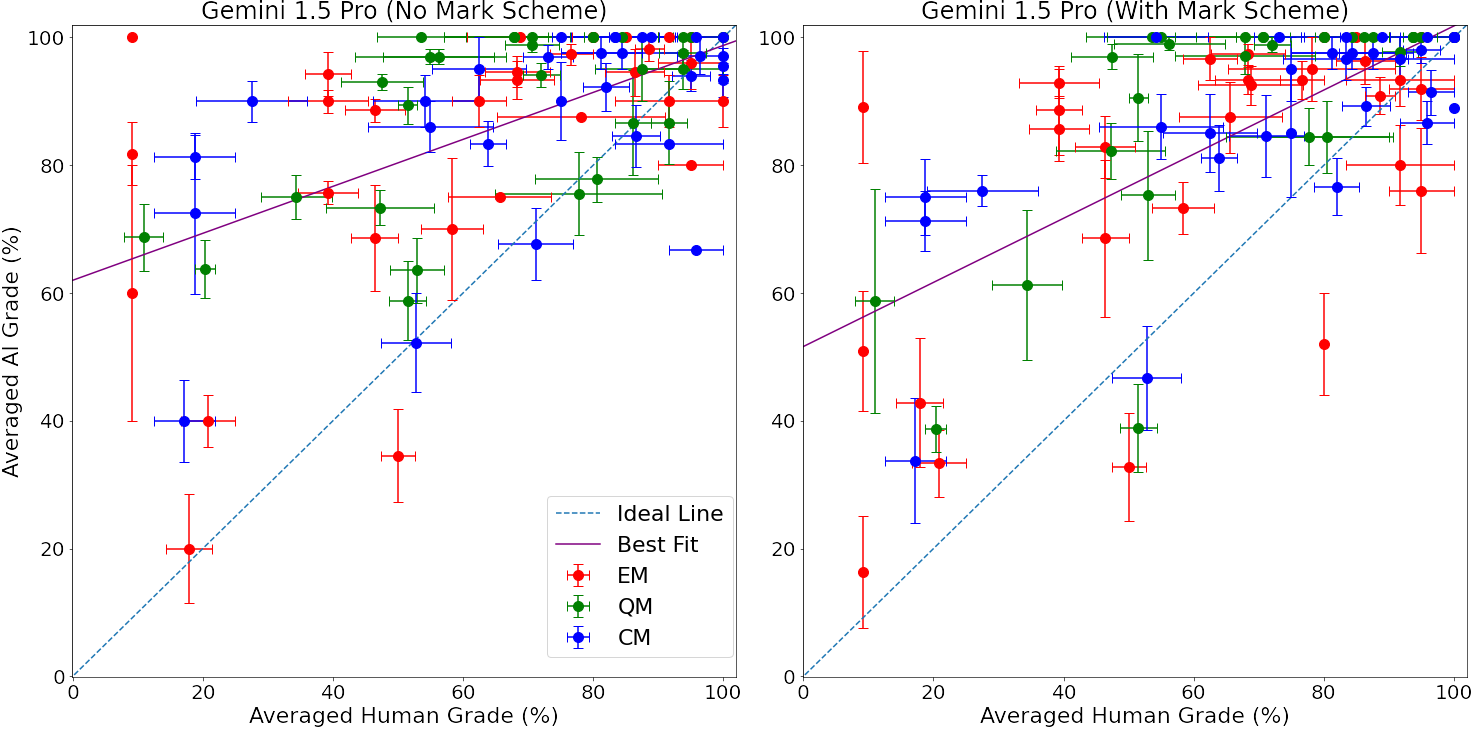}
\caption{\label{gemini_plot_withx} Regression plot of Gemini 1.5 Pro including x-errorbars. (Same as Fig. \ref{gemini_plot}).}
\end{figure*}

\begin{figure*}[hbt!]
\includegraphics[width=1\textwidth]{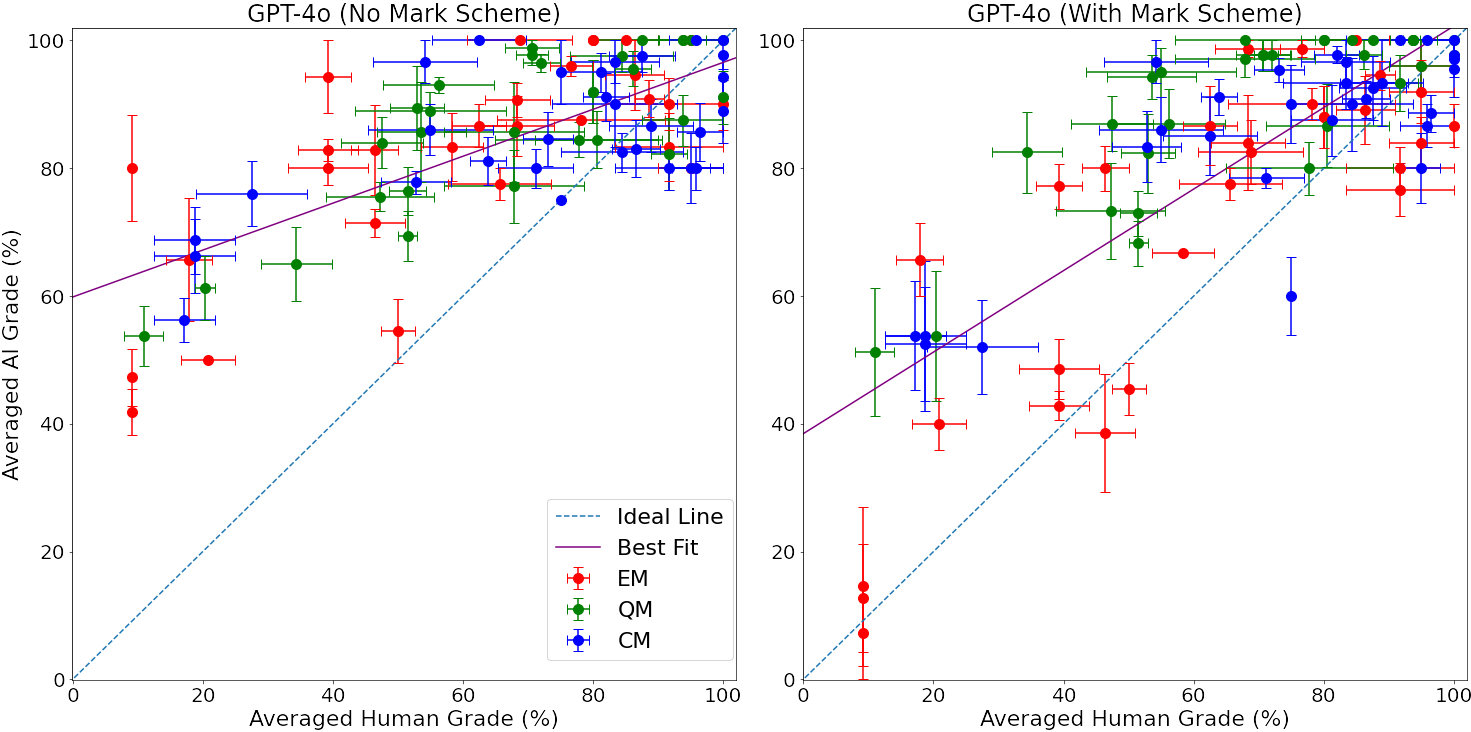}
\caption{\label{gpt4o_plot_withx} Regression plot of GPT-4o including x-errorbars. (Same as Fig. \ref{gpt4o_plot}). }
\end{figure*}

\begin{figure*}[hbt!]
\includegraphics[width=1\textwidth]{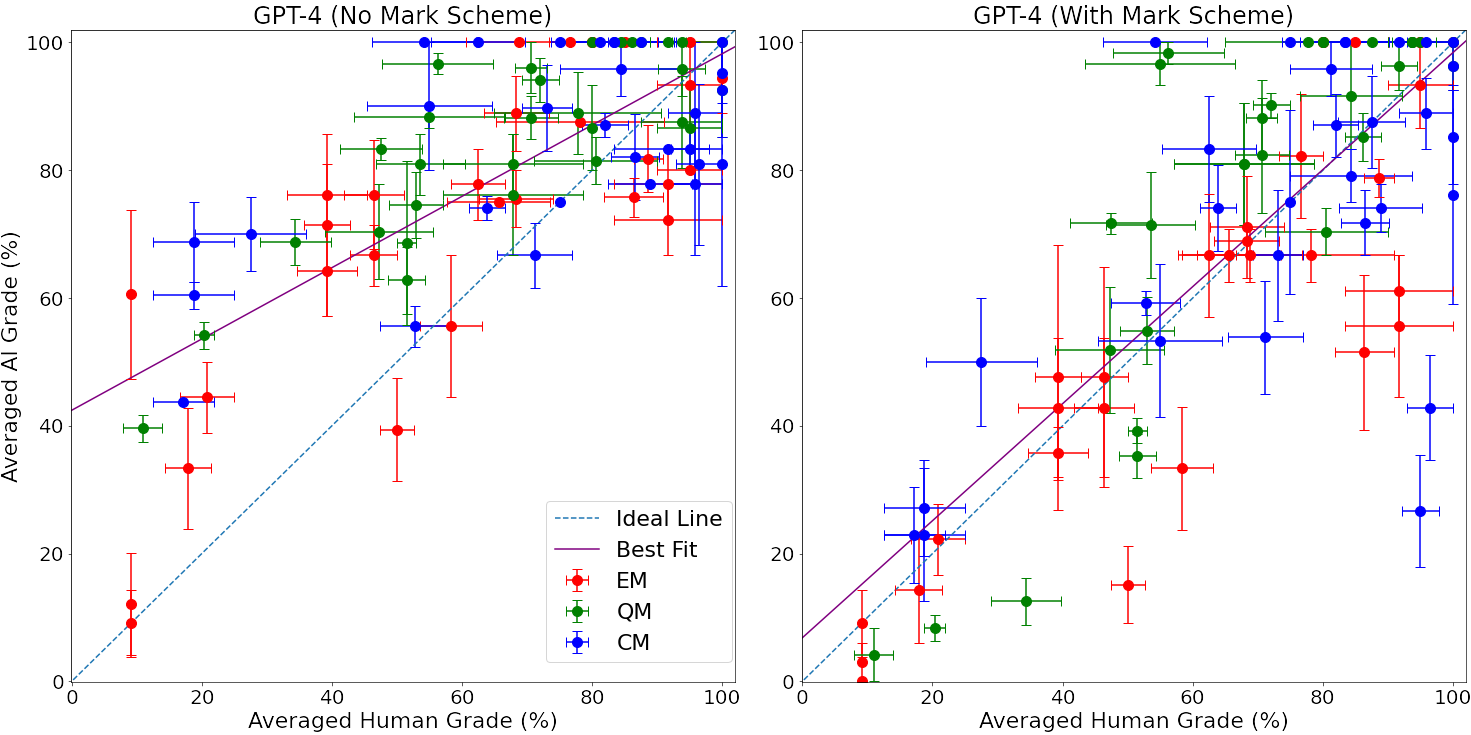}
\caption{\label{gpt4_with_x} Regression plot of GPT-4 including x-errorbars. (Same as Fig. \ref{gpt4_plot}).  }
\end{figure*}

\begin{figure*}[hbt!]
\includegraphics[width=1\textwidth]{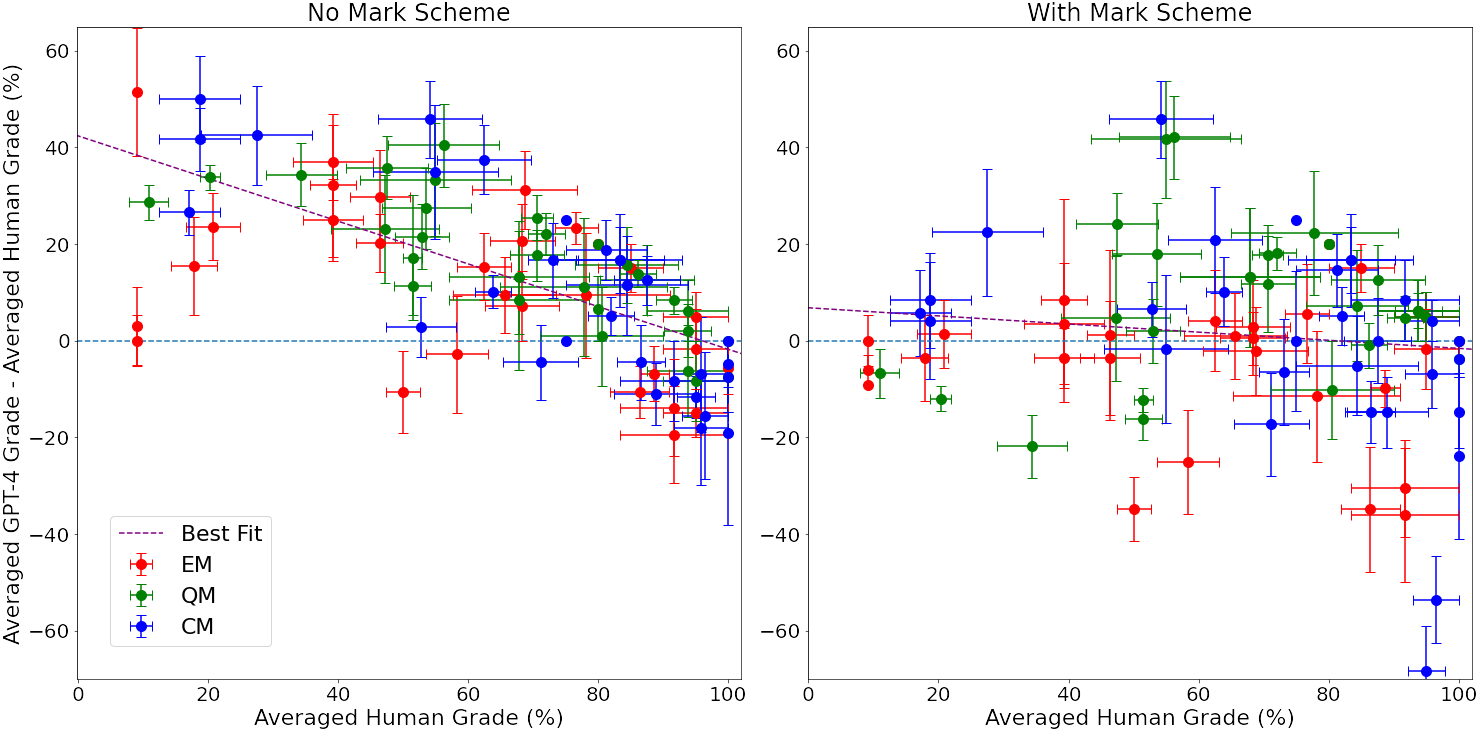}
\caption{\label{gpt4_deviation_xerr} The difference plot between human and GPT-4 grading with x-errorbars. (Same as Fig. \ref{gpt4_deviation}).}
\end{figure*}

\end{document}